\documentclass[sigconf]{acmart}
\usepackage{graphicx}
\usepackage{amsthm,amsmath} 
\usepackage{bm}
\usepackage{float}
\usepackage{multirow}
\usepackage{subfigure}
\usepackage{stfloats}

\AtBeginDocument{%
  \providecommand\BibTeX{{%
    \normalfont B\kern-0.5em{\scshape i\kern-0.25em b}\kern-0.8em\TeX}}}

\copyrightyear{2022} 
\acmYear{2022} 
\setcopyright{acmcopyright}\acmConference[ICSE-NIER'22]{New Ideas and Emerging Results }{May 21--29, 2022}{Pittsburgh, PA, USA}
\acmBooktitle{New Ideas and Emerging Results (ICSE-NIER'22), May 21--29, 2022, Pittsburgh, PA, USA}
\acmPrice{15.00}
\acmDOI{10.1145/3510455.3512782}
\acmISBN{978-1-4503-9224-2/22/05}

\begin{document}

\title{Automating Staged Rollout with Reinforcement Learning}

\author{Shadow Pritchard}
\affiliation{
\institution{University of Tulsa}
\city{Tulsa}
\state{Oklahoma}
\country{USA}
}
\email{swp7196@utulsa.edu}

\author{Vidhyashree Nagaraju}
\affiliation{
\institution{University of Tulsa}
\city{Tulsa}
\state{Oklahoma}
\country{USA}
}
\email{vidhyashree-nagaraju@utulsa.edu}

\author{Lance Fiondella}
\affiliation{
\institution{University of Massachusetts Dartmouth}
\city{Dartmouth}
\state{Massachusetts}
\country{USA}
}
\email{lfiondella@umassd.edu}

%a shortened author list if we want it
\renewcommand{\shortauthors}{Pritchard, et al.}

\begin{abstract}
Staged rollout is a strategy of incrementally releasing software updates to portions of the user population in order to accelerate defect discovery without incurring catastrophic outcomes such as system wide outages. Some past studies have examined how to quantify and automate staged rollout, but stop short of simultaneously considering multiple product or process metrics explicitly. This paper demonstrates the potential to automate staged rollout with multi-objective reinforcement learning in order to dynamically balance stakeholder needs such as time to deliver new features and downtime incurred by failures due to latent defects.
\end{abstract}

%generate using http://dl.acm.org/ccs.cfm
\begin{CCSXML}
<ccs2012>
  <concept>
      <concept_id>10011007.10011074.10011099.10011102.10011103</concept_id>
      <concept_desc>Software and its engineering~Software testing and debugging</concept_desc>
      <concept_significance>500</concept_significance>
      </concept>
 </ccs2012>
\end{CCSXML}
\ccsdesc[500]{Software and its engineering~Software testing and debugging}

% \begin{CCSXML}
% <ccs2012>
%   <concept>
%       <concept_id>10011007.10011074.10011075.10011079.10011080</concept_id>
%       <concept_desc>Software and its engineering~Software design techniques</concept_desc>
%       <concept_significance>500</concept_significance>
%       </concept>
%  </ccs2012>
% \end{CCSXML}

% \ccsdesc[500]{Software and its engineering~Software design techniques}

\keywords{DevOps, Staged Rollout, Reinforcement Learning, Software Reliability}
\maketitle

\section{Introduction}\label{sec:introduction} Researchers recognize the importance of software performance engineering and application performance management~\cite{brunnert2015performance} during the development and operation phases, including  interoperability between different tools and techniques. However, it is also important to consider process and product quality metrics in an integrated manner as well as any tradeoffs imposed. For example, staged rollout~\cite{zhao2018safely} seeks to accelerate deployment of new features while preserving safety, yet this strategy of introducing an update to a subset of users may pose tradeoffs such as the time to deliver a new feature and the downtime experienced by users. Techniques to automate the staged rollout process could reduce the level of expertise required to balance these stakeholder needs.

Tarvo et al. \cite{tarvo2015canaryadvisor} defined and collected metrics on software updates, but required the user to make release decisions. Zhao et al.~\cite{zhao2018safely} introduced a framework to release software to subsets of users based on time, power, and risk-based scheduling, which rely on predetermined thresholds or time windows. The thesis of Velayutham~\cite{velayutham2021artificial} treated staged rollout as a time series problem with autoregressive integrated moving average and long short-term memory to automate decision making, but tradeoffs were not explicitly considered.

Potential reinforcement learning approaches to model staged rollout include non-stationary~\cite{khetarpal2020towards} methods, which encompass time-varying rewards. Abdallah et al.~\cite{abdallah2016addressing} introduced repeated Q-learning to update reward tables more frequently for seldom visited states in order to improve performance under non-stationary conditions. Multi-objective reinforcement learning \cite{liu2014multiobjective} attempts to balance rewards associated with multiple objectives through a weighted, linear reward \cite{krass1990contributions}. Additional methods to balance multiple objectives include geometric steering~\cite{vamplew2017steering} or hyper volume action selection~\cite{van2013hypervolume}.

This paper presents a model of the staged rollout problem as a non-stationary Markov decision process (MDP). Multi-objective Q-learning with upper confidence bound exploration is applied to solve the MDP. The approach is demonstrated on a data set from the software reliability engineering literature, illustrating reinforcement learning as a promising approach to make dynamic decisions during staged rollout.

The remainder of the paper is organized as follows: Section~\ref{sec:model} formulates a state model of the staged rollout process, discussing tradeoffs and reward function modeling. Section~\ref{sec:algorithms} summarizes Q-learning with upper confidence bound as well as naive policy enumeration. Section~\ref{sec:metrics} develops metrics to assess policies identified by reinforcement learning. Section~\ref{sec:results} presents preliminary results. Section~\ref{sec:future} describes plans for future research.

\section{Staged Rollout Modeling and Tradeoff Assessment} \label{sec:model}
This section presents a state model of the staged rollout problem that can be interpreted as a Markov decision process (MDP), where a multi-objective reinforcement learning agent makes decisions to balance two primary factors, including (i) downtime and (ii) delivery time according to stakeholder preference. For example, safety critical software industries place greater emphasis on minimizing failures, whereas many other industries prioritize minimizing delivery time to get a product to market quickly.

\subsection{Staged Rollout  Model}\label{subsec:non-stationary}
Staged rollout of software~\cite{zhao2018safely} has been recognized as a strategy to field new functionality on an ongoing basis without incurring failures that induce system outages, widespread unavailability of services, economic losses, and user dissatisfaction. The rationale for staged rollout is to publish an updated software possessing new functionality for use by a subset of the user base to avoid the major problems described above, but also to accelerate the discovery of defects. The development team then attempts to correct the source of the problem and begins the process of staged rollout anew.

\begin{figure}[H]
    \centering
    \includegraphics[width=\linewidth]{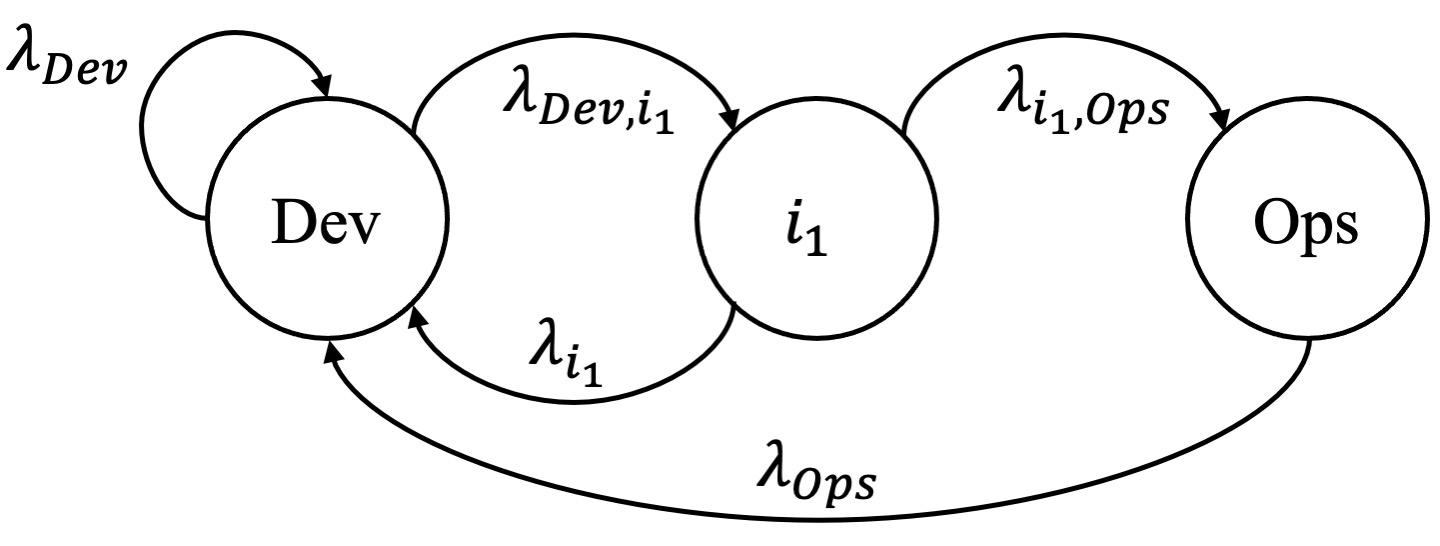}
    \caption{State diagram of
staged rollout}
    \label{fig:state_transition}
\end{figure}

Figure~\ref{fig:state_transition} shows a simple state diagram of a staged rollout process.
\noindent State $Dev$ represents the development state, where software is tested by an internal team. An elementary model of the traditional approach to software updates simply transitions to the $Ops$ state, once the software is deemed satisfactory with respect to functional requirements, reliability, and other desired attributes, where the $Ops$ state exposes the software to the entire base of $n_{Ops}$ users. Staged rollout, instead, transitions from the $Dev$ state to state $i_1$, which represents the first stage of staged rollout, where the software is published for use by $p_{i_1}$ percent or $n_{i_1}=p_{i_1}\times n_{Ops}$ of the user base. Multiple stages of staged rollout between development and full deployment are also possible. This more general case possesses $m$ intermediate staged rollout states, where it may be reasonable to assume that each transition from $i_j$ to $i_{j+1}$ increases the fraction of the user base exposed to the software such that $p_{i_j}<p_{i_{j+1}}$ and $n_{i_j}<n_{i_{j+1}}$. Failure in any state transitions to the $Dev$ state, where root cause analysis and defect removal are attempted.

The state model described in Figure \ref{fig:state_transition} enables explicit consideration of the tradeoffs between downtime and delivery time. Downtime is determined by the state in which the failure occurs and is proportional to the fraction of the user base $n_{i_j}$ and mean time to resolve ($MTTR$). Thus, failure in the $j^{th}$ state of staged rollout ($i_j$) contributes less to downtime than failure in the state $i_{j+1}$. However, the defect exposure rate in the $j^{th}$ state of staged rollout is also less than the defect exposure rate in the $(j+1)^{th}$ state, meaning that the downtime experienced, and the time required to discover and remove all defects are modeled as competing constraints. Thus, minimizing downtime by remaining in the $Dev$ state until all defects have been detected and removed will likely delay delivery time. Similarly, unrestrained transition to the Ops state immediately after each defect is resolved is likely to exacerbate downtime. Therefore, it may not be possible to simultaneously minimize downtime and delivery time, posing a multi-objective problem.  Moreover, organizations implementing staged rollout may express different levels of tolerance for these undesirable outcomes. Subsequently, it is unlikely that a single optimal policy or "one size fits all" approach to staged rollout exists. Instead, it is necessary to select transition times $t_{j,j+1}=\frac{1}{\lambda_{j,j+1}}$ that balance downtime and delivery time in a manner that is satisfactory to the customer. Intuitively, a high defect discovery rate in the $Dev$ state is likely to indicate that additional defects will be uncovered. Hence, staged rollout should not be performed because it would risk greater downtime. Therefore, the problem is to select numerical values of transition rates $\lambda_{j,j+1}=\frac{1}{t_{j,j+1}}$ that achieve the desired balance between downtime and delivery time. In the case where these rates vary as a function of time, the problem is said to be non-stationary.

Since the transitions $\lambda_{Dev,i_1}$ and $\lambda_{i_1,Ops}$ of Figure~\ref{fig:state_transition} are decisions, the staged rollout state diagram is a Markov decision process to which reinforcement learning can be applied in order to solve for a range of policies to take actions (make transition decisions) from a given state. Traditional software reliability models~\cite{inBookHoSRECh3} assume that the rate of defect discovery is nonhomogeneous. For example, if the rate of defect discovery is high in the early stages of testing, then the optimal action will be to stay in the $Dev$ state, where the penalty for failure is lowest. However, later in testing when fewer undiscovered defects remain, the optimal action will be to transition to state $i_1$ or $Ops$ in order to reduce delivery time. Therefore, the model is non-stationary because the rate at which defects are discovered changes as testing progresses. Thus, this change in the defect detection rate will also cause the reward to change and the optimal policy for a given state will change over time.

\subsection{Process Tradeoffs and Reward Modeling} \label{sec:methods}
This section defines quantitative metrics for delivery time and downtime in terms of the state model shown in Figure~\ref{fig:state_transition} as well as how these metrics are incorporated into a multi-objective reward model suitable for application of reinforcement learning.

\subsubsection{Delivery Time} \label{subsubsec:delivery time}
To model the impact of staged rollout on delivery time, we assume that one unit of time in the $Dev$ state advances the timeline by one unit, whereas time in the
staged rollout and $Ops$ state accelerate the rate at which time advances proportional to the percentage of the user base. Therefore, staged rollout may be regarded as a modern form of accelerated life testing~\cite{nelson2009accelerated} for software. For example, if the complete user base is composed of $n_{Ops} = 10,000$ users and staged rollout exposes new functionality to $p_{i_1} = 0.1$ or $10\%$ of the user base, then $n_{i_1} = 1000$. Similarly, if $n_{Dev} = 50$, then a simple method to compute the acceleration factor in each state of staged rollout is the ratio between the number of users in a state over the baseline in the $Dev$ state such that the acceleration factor in the staged rollout and $Ops$ states are $\frac{n_{i_1}}{n_{Dev}} = 20$ and $\frac{n_{Ops}}{n_{Dev}} = 200$ respectively. This simplifying assumption can be improved, since testers are familiar with the functionality and intentionally stress the program to expose defects. Modeling these ratios is a research question that requires staged rollout data. Nevertheless, the simplifying assumptions made here enable a quantitative framework upon which to improve.

The preliminary assumption of linear acceleration factors described above provides a concrete starting point to measure the cumulative time required to reach a release decision. Specifically, delivery time may be
defined as the time to discover and eliminate defects associated with a test set plus the time to transition from the $Dev$ to $Ops$ state ($t_{Dev,i_1}+t_{t_1,Ops}$) because no additional defects are discovered under the simplifying assumption that the final defect is resolved immediately. Modeling advances that explicitly consider the time between defect discovery and resolution \cite{Lo2006,nafreen2020connecting} can further enhance the realism of the staged rollout deployment model.

\subsubsection{Downtime} \label{subsubsec:downtime}
To model the impact of staged rollout on downtime, we assume that failure in the $Dev$ state does not incur downtime, since only internal testing is performed at this stage. However, downtime incurred in the staged rollout and $Ops$ states are proportional to the fraction of the user base multiplied by the mean time to resolve defects such that the accumulated downtime increases by  $p_{i_1} * MTTR$ or $p_{Ops}\times MTTR$. This simplifying assumption may be conservative, since not all users exposed to the functionality will necessarily experience the failure. Similar to delivery time, the downtime experienced must be modeled from staged rollout data and has important implications for identifying an optimal deployment policy for transition times, since conservative assumptions may unnecessarily increase delivery time. Thus, our preliminary model expresses the total downtime as the weighted sum $MTTR \times \sum_{i=1}^{n} p_{s(i)}$, where $s(i)$ denotes the state in which the $i^{th}$ failure occurs.

\subsubsection{Reward Modeling}
To model staged rollout as a reinforcement learning problem, this section defines the reward function according to techniques from multi-objective reinforcement learning \cite{liu2014multiobjective}. Specifically, a linear combination of the delivery and down time measures defined in Sections~\ref{subsubsec:delivery time} and \ref{subsubsec:downtime} at each time step. To facilitate the selection of weights in an intuitive manner, our ongoing research seeks to specify constrained optimization problems to natural forms such as (i) minimize delivery time while limiting downtime to a specified constraint or (ii) minimizing downtime while limiting delivery time to a specified constraint.
%Place explanation of normalizaiton here. Incorporate language from below and remove what you don't use. 
%\textcolor{red}{normalizing the rewards to be within a set interval assists in the design process and the effectiveness of some algorithms. Therefore, a term is added to normalize and change the sign of the reward. and the reward is on the interval $(-1,0)$.}

\section{Algorithms} \label{sec:algorithms}
This section describes the Q-learning~\cite{watkins1989learning} algorithm to estimate the rewards of state-action pairs, the upper confidence bound exploration strategy, and a naive policy enumeration approach to establish a baseline with which to compare the performance of reinforcement learning.

\subsection{Q-Learning} \label{subsec:qlearning}
Reinforcement learning \cite{sutton2018} employs a Markov decision process to estimate the reward of each state-action pair denoted $E[r(s,a)]$, which determines a policy $\pi(s)$ that informs the best action to take in any given state. Q-learning~\cite{watkins1989learning} is a model free algorithm to estimate the reward of a state-action pair according to the following update equation
\begin{equation}\label{equation:q}
    Q(s_t,a) = Q(s_t,a) + \alpha ( r + \gamma \max_{a^\prime \in \mathbb{A}} Q(s_{t+1},a^\prime) - Q(s_t, a) ) 
\end{equation}
\noindent where $s_t$ is the state at time step $t$, $a$ the action taken, $\alpha\in(0,1)$ the learning rate, $r$ the reward, $\gamma\in(0,1)$ the time discount, and $a^\prime$ the action in the next state ($t+1$) chosen from all available actions ($\mathbb{A}$). A time discount close to one expresses a willingness to defer rewards farther into the future, whereas small values of gamma increase preference for immediate rewards. The learning rate $\alpha$ is a multiplier which determines how much emphasis to place on the most recently observed reward when updating the reward estimate of a state-action pair.

\subsection{Upper Confidence Bound (UCB)}\label{subsec:UCB}
The upper confidence bound exploration strategy combines the $Q$-value with an adaptation from statistical upper confidence limits to choose the next action in a given state.
\begin{equation}\label{eq:UCB}
    a^\prime = \max_{a \in \mathbb{A}} Q(s,a) + c \times \sqrt{\log \left(\frac{t+1}{n(s,a)}\right) }
\end{equation}
\noindent where $c>0$ is a scaling constant, $t$ is the present time step, and $n$ is the number of times that a state-action pair has been selected previously.  Thus, the UCB exploration strategy selects an action according to the estimated reward as well as the uncertainty embodied in the finite sample size associated with state-action pairs. The upper confidence term is larger for smaller values of $n$, encouraging exploration of less frequently taken state-action pairs. Small values of $c$ tend toward exploitation of the action with the highest $Q$-value, whereas large values of $c$ encourage exploration of the least frequently taken actions.

\subsection{Naive Policy Enumeration} \label{subsec:naive}
Naive policy enumeration does not use reinforcement learning, is not capable of being used in an online fashion, and is therefore not a competitive alternative. It is a deterministic approach based on emulation with the full data to approximate the Pareto optimal front of tradeoffs between delivery time and downtime from a range of stationary policies in order to objectively compare the performance of a reinforcement learning approach  in terms of its distance from optimality. Specifically, state transition policy vector $\bm{\lambda}=\langle \lambda_{Dev,i_1},\lambda_{i_1,Ops}\rangle$ for Figure~\ref{fig:state_transition} is applied with a range of deterministic values such as the cross product of policies, where $\lambda_{Dev,i_1}$ and $\lambda_{i_1,Ops}$ respectively denote the time to spend in the $Dev$ and staged rollout state without failure before transitioning to the staged rollout or $Ops$ state. Thus, naive enumeration applies a range of deterministic policy vectors on the same data set and computes a tuple including the downtime and delivery time for each policy vector. A plot consisting of only the tuples that correspond to non-dominated policies (lowest delivery time for a fixed downtime or lowest downtime for a fixed delivery time) constitute the Pareto optimal front for comparison with reinforcement learning strategies.

\section{Metrics} \label{sec:metrics}
To impose greater rigor and promote objective comparison of reinforcement learning approaches, we define two quantitative metrics, including \textit{range}, which measures the flexibility of an approach, and \textit{average suboptimality}, which may be regarded as a performance-oriented metric. Both of these metrics are computed relative to a baseline Pareto optimal curve determined from naive policy enumeration. The discussion that follows introduces these metrics, provides a mathematical formula, and explains what it quantifies as well as how the metric supports objective comparison.

\subsection{Range}\label{subsec:range}
The \textit{range} of an approach ($a$) with respect to an objective ($o$) is
\begin{equation}\label{equation:range}
    range_{a,o}=\frac{\max{a,o}-\min{a,o}}{\max{naive,o}-\min{naive,o}}. 
\end{equation}
\noindent Thus, Equation~(\ref{equation:range}) is the range of values an approach achieves with respect to an objective divided by the range of the naive approach, which is treated as a baseline for the sake of normalization. The range of the naive approach is therefore $1.0$. Approaches that attain a range greater than $1.0$ for an objective produce a wider Pareto front than the naive approach, which directly impacts policy selection, since it will not be possible to identify a Pareto optimal policy below the minimum or above the maximum of a specific objective with an approach.

\subsection{Suboptimality}\label{subsec:subopt}
The \textit{suboptimality} of a specific policy obtained by an approach with respect to an objective is
\begin{equation}\label{equation:subopt}
    subopt_{p,a,o}=\frac{o_{p,a}}{o_{p,naive}}
\end{equation}
\noindent where $o_{p,a}$ is the value of the objective achieved by policy $p$ obtained with approach $a$ and $o_{p,naive}$ is the value of the objective achieved by a policy determined from the naive approach. Thus, Equation~(\ref{equation:subopt}) is simply the ratio of the value achieved by an approach for a specified objective divided by an equivalent value for the naive approach when all other objectives are held constant. In other words, we divide the value associated with a point on the Pareto curve of an approach by an equivalent point on the curve for the naive approach. Often, there is no naive policy that possesses the exact same value as the point for the policy under consideration. We therefore approximate suboptimality by linearly interpolating two points on the naive Pareto curve in order to obtain a precisely matching value. The \textit{average suboptimality} ($E[subopt_{p,a,o}]$) is a summary statistic, which computes the supoptimality for each point (individual policy) of a specified approach and objective and then calculates the average of these values.

\section{Results} \label{sec:results}
To demonstrate the potential to automate staged rollout with reinforcement learning, the SYS1 data set~\cite{lyu1996handbook}, which documents $n=136$ times at which unique defects were detected during approximately 25 hours of testing was employed. The weight on delivery time ($w_0$) was varied in the interval $(0,1)$, while the remaining weight was placed on downtime such that $w_1=1-w_0$. For each combination of weights, Q-learning with the upper confidence bound exploration strategy given in Equation~(\ref{eq:UCB}) was applied with parameters $c=0.15$, learning rate $\alpha=0.15$, and time discount $\gamma = 0.999999$. It should be noted that Q-learning with the UCB strategy learned as progress was made along the testing timeline of the SYS1 data set. Therefore, this method did not require separate data sets for training. Naive policy enumeration was performed with pairs of values from the cross product of $t_{Dev,i_1}=\{1,100,200,\dots, 10000\}$ and $t_{i_1,Ops}=\{1,100,200,\dots, 10000\}$. For the sake of illustration, parameters of the staged rollout model were set to $MTTR = 10$, $n_{Dev} = 50$, $n_i = 1000$, and $n_{Ops} = 10000$. 

\begin{figure}[!h]
    \centering
    \includegraphics[width=\linewidth]{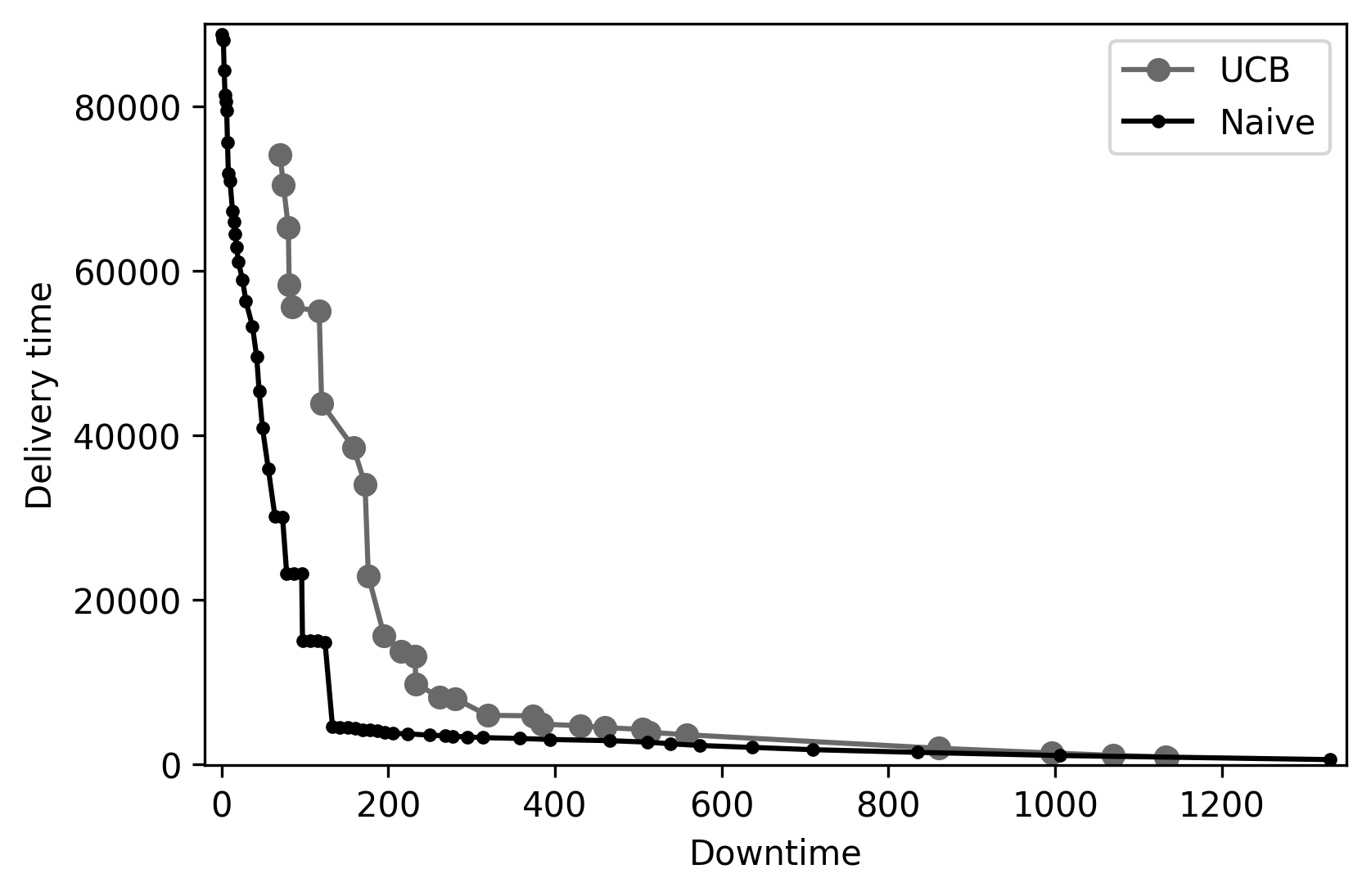}
    \caption{Tradeoff between downtime and delivery time on SYS1 data}
    \Description[]{}
    \label{fig:sys1_results}
\end{figure}

Figure~\ref{fig:sys1_results} compares the range of policies (points) identified by UCB as well as those determined by naive policy enumeration, which can only serve as a baseline because it is deterministic, unlike reinforcement learning, which makes decisions as data becomes available. Figure~\ref{fig:sys1_results} indicates that UCB policies with low delivery times and downtime greater than $300$ are competitive with naive enumeration. However, there is a visible gap between the downtime achieved by UCB compared to naive enumeration for downtimes less than $300$. One possible explanation is that UCB must perform some amount of exploration. While parameter tuning could certainly reduce this gap, a more important enhancement will be to incorporate software engineering specific testing factors that reinforcement learning can use to drive staged rollout decisions. Moreover, one may regard the results obtained by naive policy enumeration as ``lucky'', since they simply specify fixed values for times to wait before transitioning between states when no failure is experienced. The average suboptimality metric given in Equation~(\ref{equation:subopt}) summarizes this gap as a single number. For example, the average suboptimality of UCB with respect to downtime and delivery time were $2.79$ and $2.72$, indicating the potential margin for improvement.

%illustrating  the tradeoff between downtime and delivery time of the the non-stationary reinforcement learning approach with naive enumeration as a baseline.

Figure~\ref{fig:sys1_results} also shows that the best policy to minimize downtime that UCB could identify was approximately $70$ (upper left most grey dot on Pareto curve). Thus, naive enumeration was able to achieve less downtime at the expense of higher delivery times. The range metric given in Equation~(\ref{equation:range}) summarizes the width of the UCB policies relative to naive enumeration, which were $0.80$ and $0.83$ for downtime and delivery time respectively, indicating that policies identified by UCB were approximately $80\%$ as wide as the corresponding naive policies.

\section{Future Plans}\label{sec:future}
Our preliminary results obtained for Q-learning combined with the upper confidence bound approach suggest that reinforcement learning is a viable approach to automate staged rollout. Our ongoing work will consider alternative state of the art reinforcement learning methods to match or exceed naive policies and provide a greater range of flexibility. Techniques to specify a desired balance of objectives as a constrained optimization problem will provide a layer of abstraction to automatically select and apply a suitable policy on the Pareto front. Toward this end, simple and efficient techniques such as binary search can be used to identify a policy that best matches stakeholder needs. 

\section*{Acknowledgment}
This material is based upon work supported by the National Science Foundation under Grant Number 1749635 and the Office of Naval Research under Award Number N00014-22-1-2012. Any opinions, findings, and conclusions or recommendations expressed in this material are those of the authors and do not necessarily reflect the views of the National Science Foundation or Office of Naval Research.

% \textcolor{blue}{To summarize, we observed that the greedy approach provides a flexible spectrum of policies to balance stakeholder preferences with respect to downtime and delivery time, but that the stationary approach produces higher quality policies that more closely approximately naively enumerated policies. However, the quality of policies identified by the stationary and all other approach must be evaluated with testing data not employed during training.}

% \noindent Using artificially generated data results in the same limitations and successes as with the real data. UCB is not able to find all of the policies that naive enumeration can, but the range is much better this time, with policies along $95\%$ of the naive baseline for both downtime and delivery time. 

% While the range of policies improved with the generated data, the error is much worse. The average suboptimality for downtime is $6.1$ for UCB. Most of this is the result of the poor policies in the downtime interval of $(200,300)$. While naive enumeration is able to find good policies around downtime $=200$, UCB is not able to. Once downtime reaches $300$, the results for the two methods are much closer. 

\bibliographystyle{ACM-Reference-Format}
\bibliography{main}

\end{document}